\documentclass[pra,twocolumn,groupedaddress,shownopacs,amssymb]{revtex4-2}
\usepackage{graphicx,psfrag,amsmath,amssymb}
\usepackage[usenames, dvipsnames]{color}
\usepackage{braket}
\usepackage{amsmath}
\usepackage{amssymb}
\usepackage{gensymb}
\usepackage{makeidx}
\usepackage{mathtools}
 
\usepackage[normalem]{ulem}
\usepackage{float}
\usepackage[T1]{fontenc} 
\usepackage{balance}
\usepackage{flushend}
\usepackage{lineno}
\usepackage{epsfig}
\usepackage{subfigure}
\usepackage{mathtools}
\usepackage{setspace}
\usepackage{bm}
\usepackage{ulem}
\usepackage{times}
\usepackage{enumitem}
\usepackage{xcolor}
\usepackage{multirow}
\usepackage{soul}
\usepackage{dcolumn}
\usepackage{balance}
\usepackage{epstopdf}
\usepackage{braket}
\usepackage{color}
\usepackage{diagbox}
\usepackage[utf8]{inputenc}
\usepackage{hyperref}
\hypersetup{
    colorlinks=true,
    linkcolor=Blue,
    citecolor=Blue,
    filecolor=Blue,
    urlcolor=Blue,
    }
\begin{document}
\title{Collective modes of two-species Bose-Einstein condensates in a Josephson junction barrier}
\author{Harsimranjit Kaur}
\affiliation{Department of Physics, Central University of Rajasthan, Ajmer - 305817, India}
\author{Kuldeep Suthar}
\email[Corresponding author. ]{kuldeep.suthar@curaj.ac.in}
\affiliation{Department of Physics, Central University of Rajasthan, Ajmer - 305817, India}

\date{\today}
\begin{abstract}
The ultracold atoms are an ideal platform to implement atomtronics and Josephson junctions analogous to superconducting circuits. The collective modes of a Bose gas split by a potential barrier have been known. However, the role of barriers on the collective excitation spectra of ultracold atomic mixtures has not been examined. Here, we examine the low-lying collective modes of (an)harmonically trapped  
quasi-one-dimensional Bose-Einstein condensates in a Josephson barrier by employing the variational approach and Bogoliubov theory. We first show that the anharmonicity of the external potential leads to an increase in the critical barrier strength of mode softening in a single-species condensate. The Josephson barrier drives the softening of in-phase and out-of-phase dipole modes of two-species Bose-Einstein condensates, and consequently leads to two additional zero-energy Goldstone modes in the miscible phase, in agreement with the variational approach. Furthermore, the sandwich immiscible state results in an additional Goldstone mode due to the barrier, in contrast to the spatially symmetry-broken side-by-side profile. Our results unveil the distinct collective response of the Josephson barrier in binary mixtures owing to interspecies atomic correlations.
\end{abstract}
\maketitle

\section{Introduction}\label{Introduction}
 The ultracold atomic gases provide a novel and flexible platform to realize the quantum phenomenon of matter-wave coherence, quantum tunneling, entanglement, and atomtronics quantum circuits in many-body
 systems~\cite{dalfovo_99,luick_20,amico_21,amico_22,zhu_23,mukhopadhyay_24}. Although the initial observations of the Josephson effect~\cite{josephson_62} between two macroscopic coherent states were made in superconductors~\cite{barone_82} and superfluids~\cite{davis_02}, the realizations of Bose-Einstein condensates (BECs) have prompted the investigations of such quantum phenomena, as these systems possess precise control over external potential and dimensionality, tunability of interatomic interaction strengths through Feshbach resonance, and are free from imperfections~\cite{javanainen_86}. In 2005, the bosonic Josephson junction of weakly coupled Bose-Einstein condensates and the resulting quantum tunneling through a barrier was first implemented in quantum gas experiment~\cite{albiez_05}. The tunneling phenomenon has many technological applications in semiconductor diodes~\cite{esaki_60}, scanning tunneling microscopy~\cite{binnig_00}, and SQUIDs~\cite{ryu_13,jendrzejewski_14,ryu_20}. The height of the junction barrier can lead to interference between the internal and mutual collective motion of coupled condensates. Consequently, the collective excitation spectrum and density profiles are determined by the effective trapping potential and atomic interaction strengths. 

One of the interesting features of matter waves is the breaking of ${\mathbb{Z}}_{2}$ symmetry due to the presence of a repulsive Gaussian barrier in a harmonic trapping potential, leading to a double-well potential. The symmetry breaking has a direct consequence on the mode softening, degeneracy, and mode bifurcations in the collective mode spectrum of the system. It is noteworthy that the breaking of continuous symmetry in a condensate leads to a phase-coherent ground state and a gapless excitation spectrum with zero-energy Goldstone mode~\cite{kapusta_81,yukalov_07,takahashi_15}. The collective motion of the center of mass in the trap reflects one quantum energy unit of the first excited dipole mode. Symmetric barrier fragments system in two condensates, and the excitation spectrum exhibits degeneracy of low-lying modes and softening of dipole mode~\cite{salasnich_99,danshita_05, japha_11}. The Josephson oscillation frequency corresponds to that of the first lowest excited Bogoliubov mode~\cite{burchianti_17}. The nonlinear mixing of modes in the quantum dynamics results in a self-trapping regime, where population transfer to the excited state occurs during the time evolution~\cite{garaot_18}. The critical current across the barrier potential follows Berezinskii-Kosterlitz-Thouless scaling, and the exponents agree well with the corresponding equilibrium system~\cite{singh_20}. Moreover, the population dynamics of collapse-revival~\cite{sun_09} and macroscopic quantum self-trapping~\cite{sahel_02,mazzarella_09,garcia_14}, tunneling suppression~\cite{kartashov_18}, and control of Josephson oscillations through interspecies interactions~\cite{xu_08,satija_09,syu_20} in ultracold atomic mixtures have been studied. These theoretical studies reveal a direct connection between the dynamical behavior and collective mode spectrum~\cite{momme_19}. Furthermore, a mixture of two-species condensates exhibits a phase separation phenomenon. The system can be driven to a quantum phase transition from miscible to immiscible by tuning interspecies interatomic interaction strength. Several studies have discussed the collective mode excitations and their
characterizations~\cite{morise_00,franzosi_03,roy_14,suthar_15,pal_17,liu_24}, dispersion relations across the phase transition~\cite{ticknor_14,suthar_16}, and thermal effects~\cite{roy_15,lingua_15,suthar_17,roy_17}. More recently, the effects of long-range dipolar interactions in the emergence of novel states of mixtures have been investigated~\cite{bisset_21,scheiermann_23,arazo_23}. However, the influence of the barrier height of a double-well potential on the collective modes of a binary mixture has remained unexplored.  

In the present work, we first examine the evolution of collective modes with barrier height for a single-species condensate and the effects of anharmonicity using the variational approach and Hartree-Fock-Bogoliubov theory with Popov (HFB-Popov) approximation. As a system of two-species Bose-Einstein condensates (TBECs) can be mixed or demixed depending on the interaction strengths, we further unveil the effects of barrier height on the miscible and immiscible state of TBECs. The quasiparticle mode evolution shows mode softening and mode bifurcations, consistent with the change in the ground-state density profiles of both species. In the miscible state, the low-lying modes reveal similar responses due to mixing properties, while in the immiscible phase, distinct behavior is related to the spatial symmetry of the ground-state configurations. Our results shed new light on the interplay between interspecies many-body correlations and barrier potential through collective excitations. 

The remaining paper is organized as follows: Section~\ref{Model Hamiltonian and Variational Approach} details the variational approach and Hatree-Fock-Bogoliubov theory for single- and two-species condensates to determine the quasiparticle mode energies as a function of the repulsive Josephson barrier. In Section~\ref{Results_and_discussions}, we present and discuss the barrier-driven quasiparticle mode evolutions for single-species condensate, and miscible and immiscible regimes of TBECs. Finally, we summarize our results in Section~\ref{Conclusions}.

\section{Theory and Methods}\label{THEORY AND METHODS}
\label{Model Hamiltonian and Variational Approach}
\subsection{Variational Approach}
\label{varr_app}
\subsubsection{\textbf{Single-species BEC}}
Consider a quasi-one-dimensional Bose-Einstein condensate that is confined in a (an)harmonic trapping potential with frequencies \( \omega_y = \omega_z = \omega_\perp \gg \omega_x \). A Josephson (Gaussian) barrier is introduced at the center of the external trap which creates a symmetrical double-well potential as an effective trapping potential \cite{japha_11,burchianti_17,singh_20,singh_24}. The potential is 
\begin{eqnarray}
 V_{\rm ext}(\mathbf{r}) &=& V_{\rm trap} + V_{\rm barr} \nonumber \\
                        &=& \frac{1}{2}m\omega_x^2 \left(x^2 + \lambda^2 y^2 + \kappa^2 z^2 + \Omega \frac{x^4}{a_{\rm osc}^2} \right) \nonumber \\ 
                        &+& U_0 \exp\left(-\frac{2x^2}{\sigma^2}\right),
\end{eqnarray}
where the $m$ is mass of the atomic species, and $\lambda=\omega_y/\omega_x$ and $\kappa=\omega_z/\omega_x$ are anisotropy parameters along $y$ and $z$-directions, respectively. With the above conditions on the trapping frequencies, the condensate wave function can be integrated along $y$ and $z$, reducing to a quasi-1D condensate. $U_{0}>0$ and $\sigma$ controls the strength and width of the Josephson junction barrier. $\Omega\ll 1$ is a dimensionless parameter that controls the strength of anharmonicity of the trap. Here, $a_{\rm osc} = \sqrt{{\hbar}/{m \omega_x}}$ is oscillator length along the $x$-direction. 

 At zero temperature, the static and dynamical properties of quasi-1D condensate are well described by a one-dimensional Gross-Pitaevskii equation (GPE)
\begin{equation}
i \hbar \frac{\partial\phi(x,t)}{\partial t} =  \left[-\frac{\hbar^2}{2m} \frac{\partial^2}{\partial x^2} + V_{\rm ext}(x) + 
  g|\phi(x,t)|^2\right]\phi(x,t),
  \label{1s_gpe}
\end{equation}
where $\phi(x,t)$ is condensate wave-function, $g=2\sqrt{\lambda\kappa}\hbar^{2}Na_{s}/m$ characterizes the strength of two-body contact interaction with $N$ being the total number of atoms and $a_{s}>0$ is the repulsive $s$-wave scattering length. Here, $a_{\rm osc}$, $\omega_{x}^{-1}$, and $\hbar\omega_{x}$ are used as the dimensional units of length, time, and energy, respectively. The equation of motion corresponding to the above equation can be restated as a variational problem corresponding to the minimization of the action related to the Lagrangian density
\begin{equation}
{\mathcal L}_{\rm 1s} = \frac{i\hbar}{2} \left( \phi^* \frac{\partial \phi}{\partial t} - \phi \frac{\partial \phi^*}{\partial t} \right) 
- \frac{\hbar^2}{2m} \left|\frac{\partial\phi}{\partial x}\right|^2 - V_{\rm ext}(x)|\phi|^2 - \frac{g}{2} |\phi|^4.
\end{equation}
To characterize the quasiparticle modes of the condensate, the extremum of the density \(\mathcal{L_{\rm 1s}}\) is obtained by 
considering Gaussian ansatz as a trial wave function, given by 
\begin{eqnarray}
    \phi(x,t) = \eta(t) \exp\left[-\frac{[x - \chi(t)]^2}{2w(t)^2} + ix\alpha(t) + ix^2 \beta(t)\right].
\end{eqnarray}
The above Gaussian distribution is centered at position $\chi$ with width $w$ evolving in time. The other real variational parameters $\eta$, $\alpha$, and $\beta$ are amplitude, velocity, and inverse square root of beam curvature radius, respectively. Using the normalization condition $\sqrt{\pi}|\eta(t)|^2 w(t)=1$, we calculate the effective Lagrangian by integrating the Lagrangian density over the entire coordinate space. It is important to note that for the variational analysis, the exponential of the Gaussian potential is expanded up to the quartic term in $x$. Applying the Euler-Lagrange equations, we then obtain the equations of motion for the center and width of the condensate, read as
\begin{subequations}
\begin{eqnarray}
\ddot{\chi}+A\chi&+&2B{\chi}^3 + 3B\chi w^2 = 0,\\
\ddot{w}+Aw&+&3B w^3 + 6B\chi^2 w = \frac{1+g'w}{w^3},
\end{eqnarray}
\label{eom_1s}
\end{subequations}
where $A=\left(1-4U_0/\sigma^2\right)$ and $B=\left(\Omega+4U_0/\sigma^4\right)$ and \(g' = {g}/{\sqrt{2\pi}}\) is effective non-linear interaction. The other variational parameters are obtained as $\beta = {\dot{w}}/{2w}$ and $\alpha = \dot{\chi} - {\chi\dot{w}}/{w}$. In the absence of a barrier potential, the above equations of motion are consistent with previous studies~\cite{li_06,xue_08}. For positive anharmonic distortion, there is only one stable equilibrium point that corresponds to the stationary state $\chi_{0}=0$, and the corresponding width follows $Aw_0 + 3\left(4U_{0}/\sigma^4\right) w^{3}_{0} = (1+g'w_{0})/w^{3}_{0}$. The expansion of the coupled equations of center and width [Eq.~\eqref{eom_1s}] around the equilibrium point and further diagonalization leads to the frequencies of low-energy quasiparticle modes given as
\begin{subequations} \label{qpener_varr_1s}
\begin{eqnarray}
    \omega_{1} &=& \left(A + 3Bw_0^2\right)^{1/2}, \\
    \omega_{2} &=& \left(A + 9Bw_0^2 + \frac{3}{w_0^4} + \frac{2g'}{w_0^3}\right)^{1/2}.
\end{eqnarray}
\end{subequations}
These equations show that the barrier parameters affect the quasiparticle modes. In the absence of quartic distortion, $\omega_1$ corresponds to the dipole oscillation that characterizes the motion of the center of mass, while $\omega_2$ is the frequency of the variation of the condensate width. Note that the lowest excited (dipole) mode frequency equals the trap frequency in a harmonic trapping potential ($\Omega=0$) according to Kohn's theorem~\cite{fetter_98}. Anharmonic distortion $\Omega>0$ $(\Omega<0)$ leads to blue- (red-) shifted mode frequencies in atomic BECs~\cite{li_06,xue_08}.

\subsubsection{\textbf{Two-species BEC}}
The static and dynamical properties  of a system of trapped two-species BECs under a repulsive barrier is well described by the coupled GPEs, which are derived from the variational principle and read as 
\begin{eqnarray}
i\hbar\frac{\partial \phi_k}{\partial t} = -\frac{\hbar^{2}}{2m_{k}} \frac{\partial^2 \phi_k}{\partial x^2} +
V^{k}_{\rm ext}(x)\phi_k + g_{kk} |\phi_k|^2\phi_k \nonumber \\ + g_{12} |\phi_{3-k}|^2 \phi_k,
\end{eqnarray} 
where $k$ is species-index, $\phi_{k}\equiv\phi_{k}(x,t)$ is wave-function of $k$th condensate species, and $V^{k}_{\rm ext}$ is external potential, which includes the quartic distortion and repulsive barrier potential experienced by $k$th species. For TBECs, the length and energy are defined with respect to the first species i.e., $a_{\rm osc} = \sqrt{\hbar/m_{1}\omega_{x(1)}}$ and 
$\hbar\omega_{x(1)}$ serve as the length and energy scales for TBECs, where $\omega_{x(1)}$ is the harmonic trapping frequency corresponding to the first species. The intra- and interspecies interaction strengths are given by $g_{kk} =2\sqrt{\lambda\kappa}\hbar^{2}Na_{kk}/m_k$  and $g_{12} = {\sqrt{\lambda\kappa}\hbar^{2}Na_{12}}/m_{12}$, with \( m_{12} = {m_1}{m_2}/(m_1+m_2) \). The condensate wave function is normalized for each of the species, and the total number of atoms is $N=N_{1}+N_{2}$. 

Extending the variational analysis of the single-species condensate, we formulate the Lagrangian density to minimize the action of the coupled GPEs. The Lagrangian density is 
\begin{eqnarray}
\mathcal{L}_{\rm 2s} &=&\sum_{k=1}^{2} \bigg[ \frac{i\hbar}{2} \left( \phi_k^* \frac{\partial \phi_k}{\partial t} - \phi_k \frac{\partial \phi_k^*}{\partial t} \right) 
- \frac{\hbar^{2}}{2m_{k}} \left|\frac{\partial\phi_k}{\partial x}\right|^2 \nonumber \\ &-& V^{k}_{\rm ext}(x) |\phi_k|^2 
- \frac{g_{kk}}{2} |\phi_k|^4 \bigg] 
- g_{12} |\phi_1|^2 |\phi_2|^2.
\end{eqnarray}
Starting from the same trial wave function used for a single species condensate, we obtain the equations that govern the motion of the center and width of the condensates. By locating the stable equilibrium points and performing a linear expansion around them, the frequencies of the low-energy quasiparticle modes are 
\begin{subequations}
\begin{eqnarray}
     \omega_{1} &=& \left(A + 3Bw_0^2 \right)^{1/2},\\ 
    \omega_{2} &=& \left(A + 3Bw_0^2 -\frac{2g_{12}}{\sqrt{2\pi}w_0^3}\right)^{1/2},\\
    \omega_{3} &=& \left(A + 9Bw_0^2 + \frac{3}{w_0^4}+\frac{g_{11}}{\sqrt{2\pi}w_0^3}-\frac{g_{12}}{2\sqrt{2\pi}w_0^3}\right)^{1/2},\\
    \omega_{4} &=& \left(A + 9Bw_0^2 + \frac{3}{w_0^4}+\frac{g_{22}}{\sqrt{2\pi}w_0^3}+\frac{2g_{12}}{2\sqrt{2\pi}w_0^3}\right)^{1/2}.
\end{eqnarray}
\label{mode_2s_varr}
\end{subequations}
Here, $\omega_{1(2)}$ represent the frequencies associated with in-phase and out-of-phase dipole-mode excitations, and $\omega_{3(4)}$ correspond to quadrupole-mode excitations. The above mode frequencies indicate that the out-of-phase excitations possess lower energy than their in-phase counterparts for repulsive interspecies interaction strengths.

\subsection{Hartree-Fock-Bogoliubov Theory}
\label{Hartree-Fock-Bogoliubov Theory}
\subsubsection{\textbf{Single-species BEC}}
In second-quantized form, the grand-canonical Hamiltonian of $N$ interacting bosons of quasi-1D Bose-Einstein condensate confined in (an)harmonic potential under a repulsive barrier is
\begin{align} 
\hat{H} = \int dx \, \hat{\Psi}^\dagger(x,t) &\left[ -\frac{\hbar^2}{2m} \frac{\partial^2}{\partial x^2} + V_{\rm ext}(x) 
          - \mu \right. \nonumber \\
&\left. \quad + \frac{g}{2} \hat{\Psi}^\dagger(x,t) \hat{\Psi}(x,t) \right] \hat{\Psi}(x,t), \label{hamiltonian}
\end{align}
where $\hat{\Psi} (\hat{\Psi}^\dagger)$ is the annihilation (creation) bosonic field operator, $\mu$ is the chemical potential. All other parameters of the Hamiltonian have been defined previously. The Heisenberg equation of motion for the bosonic operator is
\begin{align}
i\hbar \frac{\partial \hat{\Psi}(x,t)}{\partial t} = \hat{h}\hat{\Psi}(x,t) + g \hat{\Psi}^\dagger(x,t)\hat{\Psi}(x,t)\hat{\Psi}(x,t),\label{eom}
\end{align}
where $\hat{h} = \left(-\hbar^2/2m\right)\partial^2/\partial x^2 + V_{\rm ext}(x) - \mu$
is the single-particle Hamiltonian. At temperatures below the critical value \cite{dodd_98}, where a macroscopic occupation of the ground state occurs, the condensate component can be decoupled from the Bose field operator. The non-condensed atoms, or thermal cloud, correspond to fluctuations around the condensate. We can thus express \(\hat{\Psi}(x,t)\) as $\hat{\Psi} = \phi(x) + \tilde{\psi}(x,t)$,
where \(\phi(x)\) is a classical field describing the condensate state and \(\tilde{\psi}(x,t)\) accounts for the quantum (thermal) fluctuations at zero (finite) temperatures. The generalized Gross-Pitaevskii equations under the time-independent HFB-Popov approximation~\cite{griffin_96} is
\begin{equation}
 \hat{h}\phi(x) + g\left[n_c(x) + 2\tilde{n}(x)\right]~\phi(x) = 0.\label{stationaly_state_solution}
\end{equation}
Here, \( n = n_c + \tilde{n} \) is the sum of condensate and non-condensate atomic densities with $n_{c}\equiv|\phi(x)|^{2}$ and 
$\tilde{n} = \langle\tilde{\psi}^{\dagger}\tilde{\psi}\rangle$. Thus, the last term of the equation emerges due to the presence of quantum fluctuations at zero temperature [cf. the time-dependent~Eq.~(\ref{1s_gpe})]. Using Bogoliubov transformation~\cite{bogoljubov_58}, the fluctuations can be expressed as a linear combination of the excited states or quasiparticle mode excitations as
\begin{equation}
    \tilde{\psi}(x, t) = \sum_j \left[ u_j(x) \hat{\alpha}_j e^{-iE_j t/\hbar} - v_j^*(x) \hat{\alpha}_j^\dagger e^{iE_j t/\hbar} \right],\label{fluctuation_operator}
\end{equation}
where $j$ is quasiparticle mode index, $E_j$ is $j$th quasiparticle energy, $u_j$ and $v_j$ are quasiparticle amplitudes, and $\hat\alpha_{j}(\hat\alpha^{\dagger}_{j})$ is quasiparticle annihilation (creation) operator satisfying bosonic commutation relations. We  use the above Bogoliubov transformation to obtain the following coupled Bogoliubov-de Gennes (BdG) equations: 
\begin{subequations}
\begin{eqnarray}
 (\hat{h} + 2gn) u_j - g\phi^2 v_j = E_j u_j,\;\;\;\;\;\;\\ -(\hat{h} + 2gn) v_j + g \phi^{*2} u_j = E_j v_j.\;\;\;\;\;\;\;\;\;
\end{eqnarray}
\label{bdg1s}
\end{subequations}
These coupled equations are solved self-consistently to obtain quasiparticle mode energies and amplitudes. The non-condensate density at temperature $T$ is $\tilde{n} = \sum_j \{\left[ |u_j|^2 + |v_j|^2 \right] N_0 (E_j) + |v_j|^2\}$, where $N_{0}(E_{j}) = \left(e^{\beta E_j} - 1\right)^{-1}$, with $\beta=(k_{B}T)^{-1}$, is the Bose-distribution factor of the $j$th quasiparticle state. At zero temperature, the \( N_0(E_j)\) vanishes, leading to the non-condensate density solely due to quantum fluctuations. 

\subsubsection{\textbf{Two-species BEC}}
The grand-canonical Hamiltonian for a mixture of two interacting trapped Bose-Einstein condensates in a quasi-1D (an)harmonic potential under a repulsive barrier at the center is given by
\begin{eqnarray}
    \hat{H} &=& \int dx \sum_{k=1}^{2} \hat{\Psi}_k^\dagger (x, t) \Bigg[-\frac{\hbar^2}{2m_k} \frac{\partial^2}{\partial x^2} 
   + V^{k}_{\rm ext} (x) \nonumber \\
   &-& \mu_k  + \frac{g_{kk}}{2} 
   \hat{\Psi}_k^\dagger (x, t) \hat{\Psi}_k (x, t) 
   \Bigg] \hat{\Psi}_k (x, t)  \nonumber \\
   &+& g_{12} \int dx \, 
   \hat{\Psi}_1^\dagger (x, t) \hat{\Psi}_2^\dagger (x, t) 
   \hat{\Psi}_1 (x, t) \hat{\Psi}_2 (x, t).
\label{hamiltonian_2s}
\end{eqnarray}
Here, \( k = 1, 2 \) denotes the species index. The Heisenberg equation of motion for the Bose field operator $\hat{\Psi}_k (x)$ corresponding to binary condensate is
\begin{equation*}
i \hbar \frac{\partial}{\partial t}
\begin{pmatrix}
\hat{\Psi}_1 \\
\hat{\Psi}_2
\end{pmatrix}
=
\begin{pmatrix}
\hat{h}_1 + g_{11} \hat{\Psi}_1^\dagger \hat{\Psi}_1 & g_{12} \hat{\Psi}_1^\dagger \hat{\Psi}_2 \\
g_{12} \hat{\Psi}_2^\dagger \hat{\Psi}_1 & \hat{h}_2 + g_{22} \hat{\Psi}_2^\dagger \hat{\Psi}_2
\end{pmatrix}
\begin{pmatrix}
\hat{\Psi}_1 \\
\hat{\Psi}_2
\end{pmatrix},
\end{equation*}
where $\hat{h}_{k} = \left(-\hbar^2/2m_{k}\right)\partial^2/\partial x^2 + V^{k}_{\rm ext}(x) - \mu_{k}$ is the single-particle Hamiltonian. Using the Bogoliubov approximation, the field operators of TBEC can be written as $ \hat{\Psi}_k (x, t) = \phi_k (x) + \tilde{\psi}_k (x, t)$. The equation of motion of the fluctuation operator for the first species is
\begin{eqnarray}
    i\hbar\frac{\partial\tilde{\psi}_1}{\partial t} &=& 
\left[-\frac{\hbar^2}{2m_1} \frac{\partial^2}{\partial x^2} + V^{1}_{\rm ext} + 2g_{11} (n_{1c} + \tilde{n}_1) - \mu_1 
\right] \tilde{\psi}_1 \notag \\
&+& g_{12} |\phi_2|^2 \tilde{\psi}_1 + g_{12} \tilde{n}_2 \tilde{\psi}_1 + g_{11} (\phi^{2}_1 +\tilde{m_1})\tilde{\psi}_1^\dagger \notag \\
&+& g_{12} \phi_2^* \phi_1 \tilde{\psi}_2 + g_{12} \phi_1 \phi_2 \tilde{\psi}_2^\dagger.
\end{eqnarray}
Likewise, the equation of motion for the second species is
\begin{eqnarray}
    i \hbar \frac{\partial \tilde{\psi}_2}{\partial t} &=& 
\left[-\frac{\hbar^2}{2m_2} \frac{\partial^2}{\partial x^2} + V^{2}_{\rm ext } + 2g_{22} (n_{2c} + \tilde{n}_2) 
- \mu_2 
\right] \tilde{\psi}_2 \notag \\
&+& g_{12} |\phi_1|^2 \tilde{\psi}_2 
+ g_{12} \tilde{n}_1 \tilde{\psi}_2  + g_{22} (\phi_2^2 +\tilde{m_2})\tilde{\psi}_2^\dagger \notag \\
&+& g_{12} \phi^{*}_1 \phi_2 \tilde{\psi}_1 + g_{12} \phi_1 \phi_2 \tilde{\psi}_1^\dagger.
\end{eqnarray}
We further use the Bogoliubov transformation to write fluctuation operators in terms of quasiparticle mode excitations of two-species BECs. This leads to the coupled BdG equations~\cite{roy_14} 
\begin{subequations}
\begin{eqnarray}
 \hat{{\mathcal L}}_{1}u_{1j}-g_{11}\phi_{1}^{2}v_{1j}+g_{12}\phi_1 \left 
   (\phi_2^{*}u_{2j} -\phi_2v_{2j}\right )&=& E_{j}u_{1j},\;\;\;\;\;\;\\
    \hat{\underline{\mathcal L}}_{1}v_{1j}+g_{11}\phi_{1}^{*2}u_{1j}-g_{12}
    \phi_1^*\left (\phi_2v_{2j}-\phi_2^*u_{2j} \right ) 
     &=& E_{j}v_{1j},\;\;\;\;\;\;\\
    \hat{{\mathcal L}}_{2}u_{2j}-g_{22}\phi_{2}^{2}v_{2j}+g_{12}\phi_2\left 
    ( \phi_1^*u_{1j}-\phi_1v_{1j} \right ) &=& E_{j}u_{2j},\;\;\;\;\;\;\\
\hat{\underline{\mathcal L}}_{2}v_{2j}+g_{22}\phi_{2}^{*2}u_{2j}-g_{12} 
\phi_2^*\left ( \phi_1v_{1j}-\phi_1^*u_{1j}\right ) &=& 
E_{j}v_{2j},\;\;\;\;\;\;\;\;\;
\end{eqnarray}
\label{bdg2s}
\end{subequations}
where $\hat{{\mathcal L}}_{k}=\big(\hat{h}_k+2g_{kk}n_{k}+g_{12}n_{3-k})$ with $\hat{\underline{\cal L}}_k  = -\hat{\cal L}_k$. These equations are solved self-consistently to obtain the quasiparticle modes of TBECs. The non-condensate components, or total sum of the thermal and quantum fluctuations for each of the species are 
\begin{equation}
\tilde{n}_k = \sum_j \left[ \left( |u_{kj}|^2 + |v_{kj}|^2 \right) N_0 (E_j) + |v_{kj}|^2 \right].
\end{equation}
Until the solutions converge to the required level of accuracy, the coupled Eqs.~(\ref{bdg2s}) are solved iteratively. We diagonalize coupled equations numerically using the ZGEEV routine from the LAPACK library~\cite{anderson_99,roy_20} and compute the quasiparticle energies $E_{j}$ and amplitudes $u_{kj}$ and $v_{kj}$ of the $j$th mode for the $k$th species. Here, we consider an orthonormal harmonic eigenbasis, and the number of basis is chosen to ensure the matrix contains real eigenvalues.   

\section{Results and Discussions}\label{Results_and_discussions}
 In this section, we discuss the quasiparticle mode evolutions of single-species and binary condensate mixtures obtained by numerically solving the coupled BdG equations [Eqs.~(\ref{bdg1s}) and Eqs.~(\ref{bdg2s})]. The numerical quasiparticle mode energy evolutions are further compared with the predictions of the variational approach. We first present the evolution of the low-lying quasiparticle modes and the characteristic changes as a function of the barrier height for a single-species condensate. To this end, we consider harmonically trapped \(^{23}\text{Na}\) atomic species with the number of atoms $N = 100$ and a scattering length of \( a_s = 56 a_0 \) . The trapping frequency along the \(x\)-direction is \( \omega_x = 2 \pi \times 19 \, \text{Hz} \), while the transverse frequencies in the \(y\)- and \(z\)-directions are \( \omega_y = \omega_z = 2 \pi \times 250 \, \text{Hz}\)~\cite{andrews_97,salasnich_99}. Here, the width of the barrier in scaled unit is $\sigma = 7$.  

 
\subsection{\textbf{Barrier-induced mode evolution in (an)harmonic trap}}
\label{Mode Evolution of Single-species BECs in a double well Trap and the Effects of Anharmonic Distortion}
\begin{figure}[h]
    \includegraphics[width=\linewidth]{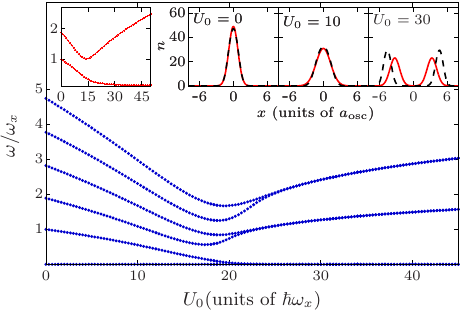}
    \caption{The change of the energies of the low-lying quasiparticle modes as a function of the barrier strength $U_{0}$ introduced to quasi-1D condensate. The excitation spectrum shown in the main  
             figure is obtained by the numerical diagonalization of BdG equations and the top left inset plot presents the two lowest excited mode frequencies obtained using analytical variational analysis. The corresponding ground-state density profiles for $U_{0} = 0, 10,$ and $30$ are shown in the inset plot, where the solid red lines are the numerical solution and 
             dashed black lines are obtained by the analytical variational ansatz. The quasiparticle frequencies are scaled to trap frequency while barrier strengths are in terms of 
             $\hbar\omega_{x}$.}
    \label{mode_1s}
\end{figure}

\begin{figure}[h]
    \includegraphics[width=\linewidth]{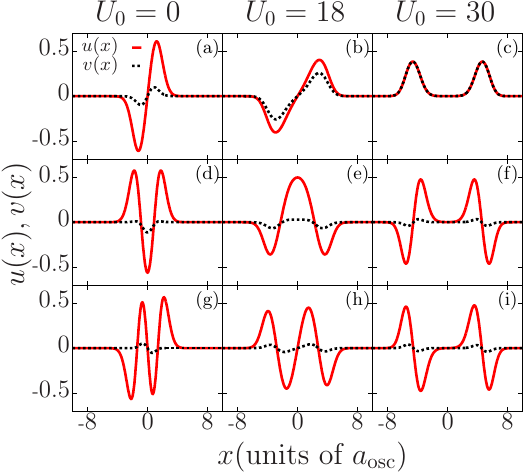}
    \caption{The evolution of quasiparticle amplitudes corresponding to Kohn mode (top row) and higher symmetric and antisymmetric modes 
             (middle and bottom rows) as barrier strength increases. At $U > U^{\rm cr}_{0}$, these low-lying modes transformed to the modes of two off-trap-centered quasi-1D BECs and thus represent an additional zero-energy mode and out-of-phase degenerate dipole modes. The shown mode amplitudes are obtained by the numerically diagonalization of BdG equations.}
    \label{qp_1s}
\end{figure}

We discuss the effects of the height of a symmetric barrier on quasiparticle energies and corresponding mode functions for a quasi-1D condensate. Fig.~\ref{mode_1s} shows the evolution of low-lying mode energies with $U_0$. This figure reproduces the collective mode energies under a double-well potential from previous studies using the multi-mode Bogoliubov theory~\cite{salasnich_99,japha_11}. Here we examine the low-lying modes with a detailed structural transformation corresponding to the change in quasiparticle energy and provide a comparison with variational analysis. The breaking of spontaneous symmetry results in a zero-energy Goldstone mode, which is apparent in Fig.~\ref{mode_1s}, as $U_{0}$ is tuned. At $U_{0}=0$, the energy of the first excited dipole (Kohn) mode is equal to the trap frequency that satisfies Kohn's theorem. The dipole mode energy further gets softened as $U_{0}$ increases and at a critical $U^{\rm cr}_{0}\approx 18.6 \hbar\omega_{x}$, it contributes to zero-energy mode leading to an additional Goldstone mode. This occurs because the repulsive barrier potential at a critical strength symmetrically bifurcates the condensate, and the system is characterized as two topologically distinct off-centered BECs having two Goldstone modes~\cite{salasnich_99}. This is evident from the evolution of ground-state density profiles with the barrier strength shown in the inset of Fig.~\ref{mode_1s}. The quasiparticle mode amplitudes of the additional mode share similar configurations as the condensate ground  state density profiles. The evolution of the dipole mode function of \text{Na} corresponding to the softening are presented in Fig.~\ref{qp_1s}(a,b,c). A dipole mode of trap-centered BEC transforms into two Goldstone modes beyond the critical value. These zero-energy modes are off-centered as determined by two isolated condensates in the ground state of the system. Thus, the transformation of the dipole mode to an additional Goldstone mode is consistent with the change in quasiparticle energies.  

Above $U^{\rm cr}_{0}$, apart from the lowest excited mode softening, the low-lying excited modes become degenerate as the system consists of two isolated condensates. The mode bifurcations and occurrence of degeneracy for the pair of excited modes (above dipole mode at $U_{0}>U^{\rm cr}_{0}$) are evident in Fig.~\ref{mode_1s}. We further analyze the structural transformation of the quadrupole mode and next excited mode with three nodes. In the absence of barrier potential, these excited modes are shown in Fig.~\ref{qp_1s}(d,g). Beyond $U^{\rm cr}_{0}$, these two higher excitations transform in two degenerate dipole modes corresponding to two isolated off-trap-centered condensates. It is worth noting that the degenerate dipole modes in the strong barrier limit are of two kinds: one belongs to in-phase while another corresponds to out-of-phase excitations.  

We next obtain the low-lying modes with barrier potential using variational analysis discussed in Section~\ref{varr_app}. We use the first two low-lying quasiparticle mode energy expressions [Eq.~\ref{qpener_varr_1s}]. The evolution of two low-lying modes obtained using variational analysis is depicted in the inset of Fig.~\ref{mode_1s}. The qualitative behaviour of the evolution of two low-lying modes agrees with the numerical results; however, the mode energies of the variational approach deviate from the numerical results for near and above critical barrier strengths~\cite{ratismith_13}. Nevertheless, the barrier-induced softening of dipole mode energy is also predicted in the analytical theory. The $U^{\rm cr}_{0}$ is determined by the chemical potential of the system~\cite{japha_11}; once the barrier strength nearly exceeds $\mu$, the additional Goldstone mode and mode degeneracy appear in the quasiparticle spectra, which follows the change in the ground-state configurations. 

\begin{figure}[h]
    \centering
    \includegraphics[width=\linewidth]{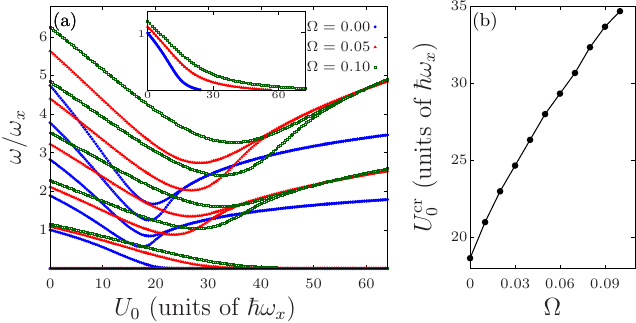}
    \caption{(a) The evolution of low-lying quasiparticle mode energies as a function of barrier strength \( U_0 \) for different 
             anharmonicity parameters $\Omega = 0.0, 0.05$, and $0.1$. The main figure shows numerical results while inset of the figure shows the mode energies obtained using the analytical variational approach. (b) The variation of critical barrier strength $U^{\rm cr}_{0}$ at which the first excited (dipole) mode gets softened as a function of the anharmonicity parameter. Here, $\Omega$ is a dimensionless parameter.}
    \label{anh_mode_freq}
\end{figure}

We further examine the effects of quartic distortion due to an anharmonic trap. The quartic potential leads to a shift in the excitation, in particular the repulsive strength $(\Omega>0)$ induces a blue-shift in the frequencies~\cite{li_06}. When the potential is perfect with no distortion, i.e., $\Omega=0$, the evolution of mode energies is discussed earlier; concerning it, we compute the mode energies for finite $\Omega$. The effects of $\Omega$ on low-lying mode frequencies are illustrated in Fig.~\ref{anh_mode_freq}. We consider $\Omega=0.05$ and $\Omega=0.1$ to understand the role of anharmonicity. As the anharmonicity strength increases, the critical barrier strength of bifurcating condensates, or introducing additional zero-energy mode, increases.  Moreover, the low-lying modes above the dipole mode become degenerate at higher $U_{0}$ (compared to the harmonic case) as the anharmonicity parameter is tuned. The qualitative behavior of the effects of $\Omega$ on dipole mode softening is also captured using variational approach, which is shown in the inset of the figure. We find that the $U^{\rm cr}_{0}$ shifts to higher values with $\Omega$ due to an increase in condensate width as a response to anharmonic distortion. With anharmonicity, the sloshing motion of the condensate requires larger barrier height to bifurcate it into two individual condensates and  thus $U^{\rm cr}_{0}$ increases. We expect that the effects of the barrier will be more prominent for atomic species with higher interatomic interaction strengths. The width of the condensate decreases with the anharmonic distortions when the barrier is absent or the barrier strength is sufficiently smaller such that the condensate does not split into two distinct condensates. This agrees with the previous study on the effects of anharmonicity~\cite{li_06}.
\subsection{\textbf{Binary condensate mixtures in (an)harmonic trap}}
\label{barr_mis}
We further examine the quasiparticle mode evolution of a mixture of two Bose-Einstein condensates. To this end, we consider a system consisting of isotopes of rubidium atomic species \(^{85}\text{Rb}\)-\(^{87}\text{Rb}\) and a mixture of two different atomic species \(^{23}\text{Na}\)-\(^{87}\text{Rb}\). The condensate mixtures exhibit a remarkable phenomenon: \textit{phase separation} when $g_{12}$ exceeds the geometric mean of intraspecies interaction strengths. However, this condition is crucially affected by the trapping potential and number of atoms \cite{trippenbach_00,yukalov_07,wen_12,wen_20,pyzh_20}. This leads to primarily two phases: miscible and immiscible phases. In the miscible phase, both species' ground state density profiles overlap while stronger interspecies interaction results in phase-separated condensate density configurations. The ground state density profile of \(^{85}\text{Rb}\)-\(^{87}\text{Rb}\) mixture in the immiscible phase is \textit{side-by-side}, i.e. both species are offset to the trap center. This is due to both species being nearly in mass and having relatively small atom numbers. On the other hand, \(\text{Na}\)-\(\text{Rb}\) shows a \textit{sandwich}-type profile where the species with heavier mass remains at the center and is flanked by the lighter mass atomic species. We consider $N_{1}=N_{2}=100$ for \(^{85}\text{Rb}\)-\(^{87}\text{Rb}\) TBEC and $N_{1}=N_{2}=1000$ for  \(\text{Na}\)-\(\text{Rb}\) TBEC. We investigate the intriguing role of interspecies correlations and the external repulsive barrier potential on collective excitations due to changes in ground-state density profiles. Hereafter, we consider \(^{85}\text{Rb}\) and \(^{23}\text{Na}\) as first species and \(^{87}\text{Rb}\) as second species for the above mentioned systems. The contact interatomic interactions are determined by the scattering lengths, here $a_{11}=99a_{0}$, $a_{22}=100a_{0}$ for the \(^{85}\text{Rb}\)-\(^{87}\text{Rb}\) system and $a_{11}=56a_{0}$, $a_{22}=100a_{0}$ for the \(\text{Na}\)-\(\text{Rb}\) system, respectively, where $a_{0}$ is the Bohr radius. The trapping frequencies are \(\omega_{x(1)} = 2 \pi \times 4.55\, \text{Hz}\) and \(\omega_{x(2)} = 2 \pi \times 3.89\, \text{Hz}\) for the \(^{85}\text{Rb}\)-\(^{87}\text{Rb}\) mixture~\cite{papp_08}. The trapping frequency ratio of condensates \( \omega_x(\text{Na}):\omega_x(\text{Rb}) = 1.1 \) \cite{wang_16}. We set the anisotropy parameters $\lambda$ and $\kappa$ of the single-species condensate to achieve a quasi-\(1D\) system. These systems allow us to examine cases with both negligible and substantial mass differences between the components, as well as to explore the role of barrier on both miscible and immiscible phases with different ground-state density configurations.

\subsubsection{Barrier-induced mode evolution in miscible phase}
\begin{figure}[h]
    \centering
    \includegraphics[width=\linewidth]{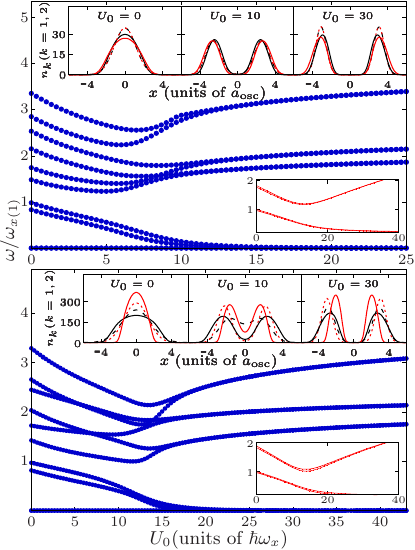}
    \caption{The evolution of low-lying quasiparticle mode energies as a function of repulsive barrier strength \( U_0 \) in the miscible phase of two-species Bose-Einstein condensates:   
             \(^{85}\mathrm{Rb}\)–\(^{87}\mathrm{Rb}\) (top panel) and \(^{23}\mathrm{Na}\)–\(^{87}\mathrm{Rb}\) (bottom panel). Barrier results in the mode softening and degeneracy of low-lying quasiparticle modes. The main figures are numerical results and insets show results obtained using the variational approach discussed in Section~\ref{varr_app} [Eq.~\ref{mode_2s_varr}]. The change in the density profiles for different $U_{0}$ values are shown in the inset plot. The numerical density of first (second) species is represented by solid 
             black (red) color lines, while the corresponding dashed lines are the variational ansatz solution.}
    \label{mode_mis}
\end{figure}

We now examine the quasiparticle mode evolution of a miscible phase of binary mixtures in the presence of a repulsive barrier. The overlapped density profiles in the miscible phase correspond to the breaking of two spontaneous symmetries, resulting in two zero-energy Nambu-Goldstone modes. The mode energies as a function of the barrier strength in the miscible phase are presented in Fig.~\ref{mode_mis} for two systems. We first discuss the modes for the miscible phase of \(^{85}\mathrm{Rb}\)-\(^{87}\mathrm{Rb}\) TBEC. For systems to be in the miscible phase, we set $a_{12}=10a_{0}$.

Contrary to a single-species case, since we have two overlapped condensates at the trap center in the miscible phase, thus the barrier potential bifurcates each of the condensates. 
The splitting of a condensate into two distinct condensates for each of the species of a binary mixture is shown in the inset of Fig.~\ref{mode_mis}.
This leads to the softening of both Kohn modes as a function of barrier strength. These two modes contribute to zero-energy modes at a $U^{\rm cr}_{0}$ value. The breaking of spatial symmetry by a barrier potential results in two additional Goldstone modes, and the system in the miscible phase possesses four Goldstone modes. Moreover, likewise in single-species condensate, the energies of low-lying modes (above Kohn mode) decrease, and around the critical strength, a pair of modes acquire degeneracy as barrier strength is ramped up. It is apparent from the degeneracy of the quadrupole mode and the next excited state, which serves as the first excited mode of separated off-centered condensates. It is worth noting that the four zero-energy modes will be the lowest energy modes above $U^{\rm cr}_{0}$. For the miscible phase of \(\text{Na}\)-\(\text{Rb}\) condensates, the qualitative features of mode softening and degeneracy with barrier remain the same. For variational analysis, we consider the same mass and frequencies, and the mode evolutions are shown in the insets of the figure. And the analysis gives a smaller difference in the mode entries for $U<U^{\rm cr}_{0}$. This is because for the analytical variational approach, the masses and trapping frequencies are considered identical for both species. The exponential of the Gaussian potential is expanded up to quartic terms (in $x$) in the variational approach. The expressions of low-lying quasiparticle mode energies deviate from the exact numerical results near and above the critical barrier strengths. Furthermore, 
identical mass assumption results in a smaller difference in mode energies. However, the results of mode softening obtained using the variational approach are qualitatively in agreement with the numerical solution of the coupled Bogoliubov-de Gennes (BdG) equations [Eqs.~(\ref{bdg2s})]. The numerical solutions of the coupled BdG equations provide an accurate description of the change in quasiparticle mode energies with barrier potential strength.

\begin{figure}[h]
    \centering
    \includegraphics[width=\linewidth]{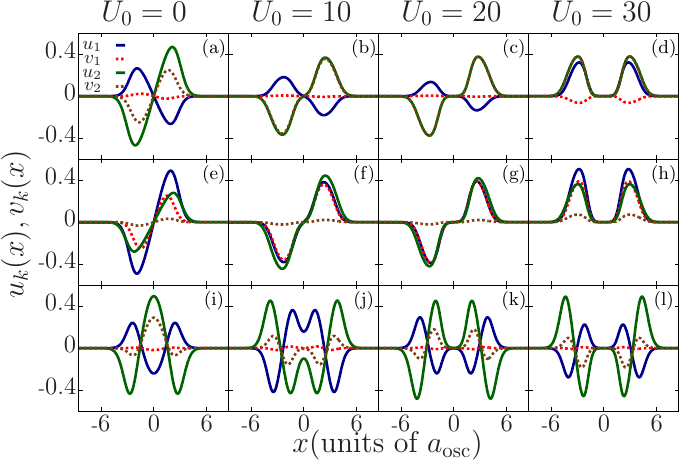}
    \caption{Shown here are the evolutions of the quasiparticle amplitudes for the miscible phase of the \(^{85}\mathrm{Rb}\)-\(^{87} \mathrm{Rb}\) two-species condensate as \(U_0\) increases from  
             zero to $30$ (in units of $\hbar\omega_{x}$). The first (second) row corresponds to the evolution of an out-of-phase (in-phase) Kohn mode, and the third row depicts the evolution of an out-of-phase quadrupole mode. The Kohn mode gets softened and transforms to two additional Goldstone modes (with already existing two Goldstone modes of spontaneous symmetry breaking of Bose-Einstein condensation of two-species condensates). Moreover, the quadrupole mode transforms to out-of-phase Kohn mode of two isolated off-centered condensates above $U^{\rm cr}_{0}$. Here, the interspecies scattering length $a_{12}=10a_{0}$. The mode amplitudes $u_{k}$ and $v_{k}$ are in units of $1/\sqrt{a_{\rm osc}}$. The mode amplitudes are scaled with appropriate factors for better visualization.}
    \label{modefn_mis}
\end{figure}

We further analyze the evolution of low-lying quasiparticle mode amplitudes or functions with barrier strengths. For brevity, we consider the evolutions of the modes of \(^{85}\mathrm{Rb}\)-\(^{87}\mathrm{Rb}\) TBEC; however, the qualitative behavior of the modes for the other system also remains similar. The structural evolution of the Kohn modes and one of the quadrupole modes with barrier strengths are shown in Fig.~\ref{modefn_mis}.
Without a barrier, the first excited mode corresponds to in-phase and out-of-phase dipole excitations in two-species condensates. In the in-phase mode, the dipole mode oscillations of species occur in consonance, i.e. $u_{1}$ and $u_{2}$ are in the same phase, while in the out-of-phase mode, the oscillations of $u_{1}$ and $u_{2}$ appear with a $\pi$ phase difference. These two first (lowest) excitation modes are shown in Fig.~\ref{modefn_mis}(a,e). The quasiparticle energy of in-phase Kohn mode remains unity (scaled with $\hbar\omega_{x}$) as per Kohn's theorem, while in general out-of-phase modes remain lower in energy. Moreover, the (out-of-phase) quadrupole mode of the system at $U_{0}$ is also shown in Fig.~\ref{modefn_mis}(i).  

As the strength of the repulsive barrier potential is tuned to a finite value, both Kohn modes tend to become flatter at the trap center; this is evident from the evolution shown in the first and second panels at $U_{0}=10$ [Fig.~\ref{modefn_mis}(b,c)] and $U_{0}=20$ [Fig.~\ref{modefn_mis}(f,g)] respectively. At $U_{0}=30$, well above the $U^{\rm cr}_{0}$, the quasiparticle amplitudes of Kohn modes correspond to those of Goldstone modes of two-species system at the same $U_{0}$ value. The structures of these modes are reminiscent of ground-state density profiles at $U_{0}=30$ (not shown here). As expected, the structural transformation of the Kohn modes with $U_{0}$ is consistent with the mode softening and energy presented in Fig.~\ref{mode_mis}. As the barrier potential bifurcates the condensate density profiles, the peak of the quadrupole mode gets a dip that approaches zero with barrier strength; the related evolution can be seen in Fig.~\ref{modefn_mis}(j,k). Finally, above $U^{\rm cr}_{0}$, the structural transformation of the quadrupole mode in the out-of-phase Kohn mode of off-centered condensates is evident in Fig.~\ref{modefn_mis}(l), which is in line with the quasiparticle mode energy evolution of \(^{85}\mathrm{Rb}\)-\(^{87}\mathrm{Rb}\) TBEC shown in Fig.~\ref{mode_mis}. Moreover, with anharmonicity $\Omega$, the $U^{\rm cr}_{0}$ of mode softening is lower for out-of-phase Kohn mode than that of in-phase mode. $U^{\rm cr}_{0}$ also depends on the mass difference of TBECs, and in particular, it is greater for larger mass difference \(\mathrm{Na}\)-\(\mathrm{Rb}\) TBEC.

\subsubsection{Barrier-induced mode evolution in immiscible phase}
\begin{figure}[h]
    \centering
    \includegraphics[width=\linewidth]{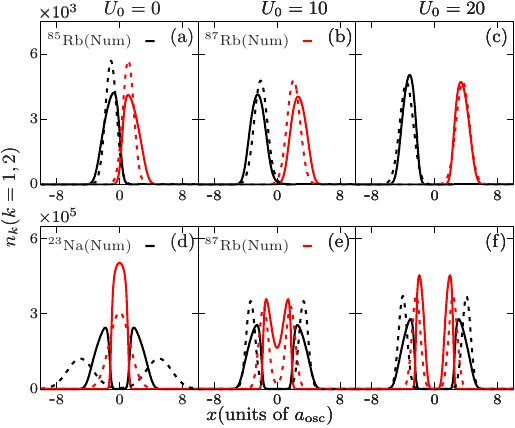}
    \caption{The geometry of the condensate density profiles in the immiscible regime as a function of $U_{0}$. The values of \(U_0\) are displayed at the top of the figures. The top panel (a,b,c)   
             shows the distributions of \(^{85}\mathrm{Rb}\)-\(^{87}\mathrm{Rb}\) TBEC with $a_{12} = 450a_{0}$ while the bottom panel (d,e,f) presents the distributions of \(^{23}\mathrm{Na}\)-\(^{87}\mathrm{Rb}\) TBEC with $a_{12} = 300a_{0}$. The density profiles are obtained from the numerical solution of the coupled GPEs. The corresponding dashed lines
             represent the solution obtained by analytical ansatz.}
    \label{den_immis}
\end{figure}

We now turn to discuss the role of barrier strengths on the collective modes in the immiscible phase. We fix $a_{12}=450a_{0}$ for \(^{85}\mathrm{Rb}\)-\(^{87}\mathrm{Rb}\) TBEC and $a_{12}=300a_{0}$ for \(\mathrm{Na}\)-\(\mathrm{Rb}\) TBEC. The two systems considered have different ground-state density configurations,  which are obtained by the numerical solution of the coupled GPEs. This leads to different responses of barriers on ground states as well as the collective excitations of the systems. The changes in the density profiles with varying barrier strengths in the immiscible phase are shown in Fig.~\ref{den_immis}. The ground-state density distributions of \(^{85}\mathrm{Rb}\)-\(^{87}\mathrm{Rb}\) and \(\mathrm{Na}\)-\(\mathrm{Rb}\) are side-by-side and sandwich geometry, respectively [Fig.~\ref{den_immis}(a,d)]. These density profiles in the absence of barrier potential agree with a previous study~\cite{roy_14}.
The repulsive Josephson barrier at the center depletes atoms from the center of the harmonic trap. This decreases the minimal overlap of the densities of the species, and repulsion causes two topologically isolated condensates [Fig.~\ref{den_immis}(b,c)]. On the other hand, the barrier bifurcates the species at the center of the trap in \(\mathrm{Na}\)-\(\mathrm{Rb}\) TBEC. The ground state density profile of the system shown in Fig.~\ref{den_immis}(f) can be assumed to be four topologically distinct condensates. Thus, introducing a repulsive barrier in the immiscible phase leads to an additional condensate. Note that the sandwich profile itself can be considered as three distinct condensates~\cite{ticknor_13,roy_14}. The collective modes (with barrier) corresponding to these immiscible states give rise to a novel excitation spectrum.   
\begin{figure}[h]
    \centering
    \includegraphics[width=\linewidth]{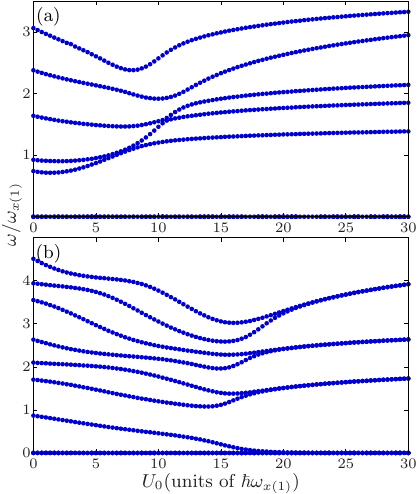}
    \caption{The evolution of low-lying mode energies as a function of \( U_0 \) in the immiscible phase of TBECs: 
             \(^{85}\mathrm{Rb}\)–\(^{87}\mathrm{Rb}\) (top panel) and \(\mathrm{Na}\)-\(\mathrm{Rb}\) (bottom panel). The mode energy evolution of TBECs with barriers is distinct owing to mass difference and consequent different phase-separated density profiles. Unlike the lower panel, the upper figure is devoid of mode softening and degeneracies.}
    \label{mode_2s_imm}
\end{figure}

\begin{figure}[h]
    \centering
    \includegraphics[width=\linewidth]{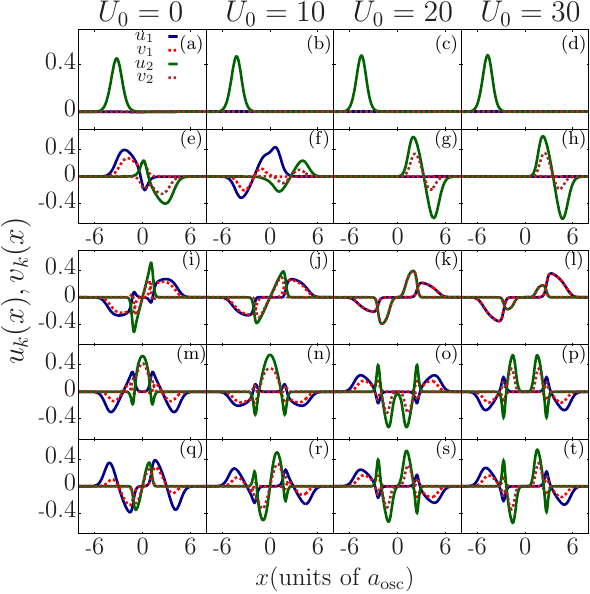}
    \caption{The evolution of quasiparticle amplitudes with barrier strengths in the immiscible regime for \(^{85}\mathrm{Rb}\)–\(^{87}\mathrm{Rb}\) binary condensates (a-h) and \(\mathrm{Na}\)-\(\mathrm{Rb}\) (i-t). Shown are the evolutions of hardened mode (first row) and dipole mode (second row) of \(^{85}\mathrm{Rb}\)–\(^{87}\mathrm{Rb}\), and first three excitations with nonzero quasiparticle energies in \(\mathrm{Na}\)-\(\mathrm{Rb}\) TBEC. Here, the mode amplitudes $u_{k}$ and $v_{k}$ are in units of $1/\sqrt{a_{\rm osc}}$.
    }
    \label{modefn_immis}
\end{figure}

Fig.~\ref{mode_2s_imm} presents the behavior of collective Bogoliubov excitations with repulsive barrier strengths in the immiscible regime of TBECs. In the immiscible phase of \(^{85}\mathrm{Rb}\)-\(^{87}\mathrm{Rb}\) TBEC, as a function of $g_{12}$, the Kohn mode energy gets softened and gets hardened at the phase separation~\cite{roy_14}. This is due to the breaking of $z$-parity symmetry, and the mode approaching zero energy regains energy. The number of Goldstone modes remains at two, owing to spontaneous symmetry breaking of condensation. As $U_{0}$ ramps up, the mode energy increases as the barrier creates a repulsion between the topologically distinct condensates. And the energy gets stabilized with $U_{0}$ once the far-apart condensates are not affected by the barrier. Hence, the quasiparticle spectra of \(^{85}\mathrm{Rb}\)-\(^{87}\mathrm{Rb}\) TBEC in an immiscible state devoid of mode softening and degeneracies; however, mode bifurcations may occur for higher excited states. We further corroborate the effect of $U_{0}$ on the mode energy of hardened and in-phase mode through the structural transformation of quasiparticle mode functions. These are shown in Fig.~\ref{modefn_immis}(a-h). The mode excitations corresponding to the hardened mode represent an off-centered peak corresponding to the second species with a phase shift concerning the density profile. The first excited mode remains unchanged with $U_{0}$ that is consistent with no change in energy. The in-phase dipole mode at $U_{0}=0$ transforms to the dipole mode corresponding to the second species. At higher energies, the decoupled modes of the first species are observed. Thus, the decoupled condensates exhibit two new modes corresponding to the lowest (finite) energy excitations of each condensate. The higher energy modes represent the dipole excitations of the condensates in the decoupled phase-separated state.   

We finally discuss the mode evolution of \(^{23}\mathrm{Na}\)-\(^{87}\mathrm{Rb}\) TBEC. The change in the quasiparticle energy is shown in Fig.~\ref{mode_2s_imm}(b). Due to significant mass differences and sandwich ground-state density distribution, the mode energy evolves distinctly from that of \(^{85}\mathrm{Rb}\)-\(^{87}\mathrm{Rb}\) TBEC. Here, the condensate at the trap center leads to an additional Goldstone mode, and the system possesses four zero-energy modes. The softening of the first excited mode towards zero energy is evident from the figure. The higher energy modes acquire mode degeneracy above a critical barrier strength. This immiscible state behaves similarly to a single-species condensate. We further analyze the evolution of the lowest mode amplitudes, which are shown in Fig.~\ref{modefn_immis}(i-t). The first panel represents the evolution of the first excited Kohn mode into (an additional) Goldstone mode as the strength $U_{0}$ increases. The absolute of the mode functions above the critical strengths of $U_{0}$ share structural similarity with the Goldstone mode that reflects the condensate density profiles. This is evident from the comparison of the absolute of Fig.~\ref{modefn_immis}(k) with Fig.~\ref{den_immis}(f). It is worth noting that the softening of the Kohn mode in double-well potential~\cite{salasnich_99} and structural transformation of Kohn modes at the phase separation have been 
studied~\cite{roy_14}. We attribute the softening of the mode to the corresponding change in the structure of mode functions.
With $U_{0}$, the other two higher energy modes [Fig.~\ref{modefn_immis}(m,q)] transform into two degenerate dipole excitations of four topologically distinct condensates [Fig.~\ref{modefn_immis}(p,t)]. In the presence of anharmonic distortion, the mode evolution qualitatively remains the same; however, like single-species and miscible phase spectra, the critical barrier strength for the emergence of an extra zero-energy mode in \(\mathrm{Na}\)-\(\mathrm{Rb}\) TBEC will increase. We explicitly obtained $U^{\rm cr}_{0}=16.66$, $20$, and $22.33$ for $\Omega=0$, $0.05$, and $0.1$, respectively. Hence, the quasiparticle mode evolution underscores the differing response of two binary condensate systems in the immiscible phase. 

\section{Conclusions}\label{Conclusions}
We have studied the collective excitation spectrum of quasi-one-dimensional single- and two-species Bose-Einstein condensates in the presence of a repulsive Josephson barrier. We first reveal the increase in critical barrier strengths of mode softening due to anharmonic distortion. Furthermore, the miscible phase of binary mixtures exhibits similar quasiparticle mode evolutions owing to overlapping condensate profiles. The barrier potential leads to softening of in-phase and out-of-phase dipole excitations into two additional zero-energy Goldstone modes, which are in agreement with the variational analysis. Moreover, the quasiparticle spectra exhibit mode degeneracies above the critical barrier strength when it exceeds the chemical potentials. While the excitation spectrum of the immiscible state is imperatively determined by the ground-state density distribution. A system with negligible mass differences and lower particle numbers does not show mode softening and degeneracy. On the other hand, the mode evolution of mixtures with two different atomic species exhibits softening of the out-of-phase Kohn mode and thus leads to an additional Goldstone mode. The present study bridges a gap in understanding the collective excitations driven by a Josephson junction and unveils the role of interspecies correlations. In relevance to the recent interest in exploring the Josephson barrier in superconducting materials and superfluid matter, the present work could inspire further theoretical studies such as the investigations of the effects of thermal fluctuations and opens a possibility of realizations of barrier-induced quasiparticle spectra in quantum gas experiments.

\begin{acknowledgments}
H.K. acknowledges the financial support from University Grant Commission (UGC), New Delhi. K.S. acknowledges support from the Science and Engineering Research Board, Department of Science and Technology, Government of India through Project No. SRG/2023/001569.
\end{acknowledgments}

\bibliography{mode_dw}{}
\bibliographystyle{apsrev4-2}
\end{document}